\documentclass[12pt]{article}
\begin{document}

\begin{center}
{\Large\bf{} Gravitational energy from a combination of a tetrad
expression and Einstein's pseudotensor}
\end{center}

\begin{center}
Lau Loi So\\
Department of Physics, Tamkang University, Tamsui 251, Taiwan
\end{center}

\begin{abstract}
The energy-momentum for a gravitating system can be considered by
the tetard teleparalle gauge current in orthonormal frames.
Whereas the Einstein pseudotensor used holonomic frames. Tetrad
expression itself gives a better result for gravitational energy
than Einstein's. Inspired by an idea of Deser, we found a
gravitational energy expression which enjoys the positive energy
property by combining the tetrad expression and the Einstein
pseudotensor, i.e., the connection coefficient has a form
appropriate to a suitable intermediate between orthonormal and
holonomic frames.
\end{abstract}

\section{Introduction}
Gravitational energy is still a fundamental problem in general
relativity.  The gravitation field energy does exist through the
physical phenomenon such as the Io heating, the tidal force for
Jupiter acts on its satellite planet Io.  Owing to the equivalence
principle, it is not meaningful to study the gravitational energy
at a point, however the quasilocal idea can get around this
difficulty. The Bel-Robinson tensor \cite{Szabados} has a
desirable property for a gravitating system quasilocally because
it preserves the positive energy for all moving observers.
Gravitational energy has been studied  using the Einstein
pseudotensor in holonomic frames for a long time \cite{MTW}.
Recently, Deser $et$ $al$. \cite{Deser} used a similar method to
calculate the Landau-Lifschitz pseudotenor and obtained the
Bel-Roboinson tensor by making a specific combination of these two
classical pseudotensors.  Lately the tetrad expression
\cite{Nester} has been evaluated in orthonormal frames and also
gave this good result. The present paper is inspired by the idea
of Deser $et$ $al$., we found a gravitational energy expression
which enjoys the positive energy property by making a specific
combination of the tetrad expression and the Einstein
pseudotensor, meaning that the connection coefficients are being
selected by a uniquely specific intermediate between orthonormal
and holonomic frames.

\section{Ingredient}
The curvature 2-form, in differential form, is
\begin{equation}
R^{\alpha}{}_{\beta}:=d\Gamma^{\alpha}{}_{\beta}
+\Gamma^{\alpha}{}_{\lambda}\wedge\Gamma^{\lambda}{}_{\beta},
\end{equation}
and the covariant derivative of the vanishing torsion is
\begin{equation}
0=D\eta_{\alpha}{}^{\beta}{}_{\mu}=d\eta_{\alpha}{}^{\beta}{}_{\mu}
+\Gamma^{\beta}{}_{\lambda}\wedge\eta_{\alpha}{}^{\lambda}{}_{\mu}
-\Gamma^{\lambda}{}_{\alpha}\wedge\eta_{\lambda}{}^{\beta}{}_{\mu}
-\Gamma^{\lambda}{}_{\mu}\wedge\eta_{\alpha}{}^{\beta}{}_{\lambda},
\end{equation}
where the dual basis
$\eta^{\alpha\cdots}:=*(\theta^{\alpha}\wedge{}\ldots)$ and
$\theta^{\alpha}$ is the co-frame.  At one point, it is well known
that one can choose Riemann normal coordinate in which the
connection coefficient satisfies
\begin{equation}
\Gamma^{\alpha}{}_{\beta\mu}(0)=0,\quad{}
-3\partial_{\nu}\Gamma^{\alpha}{}_{\beta\mu}
=R^{\alpha}{}_{\beta\mu\nu}+R^{\alpha}{}_{\mu\beta\nu}.
\end{equation}
Similarly, one can choose orthonormal frames \cite{Nester} such
that
\begin{equation}
\Gamma^{\alpha}{}_{\beta\mu}(0)=0,\quad{}
2\partial_{\nu}\Gamma^{\alpha}{}_{\beta\mu}
=R^{\alpha}{}_{\beta\nu\mu}.
\end{equation}
Define the Bel-Robinson tensor $B_{\alpha\beta\mu\nu}$ and tensor
$S_{\alpha\beta\mu\nu}$ in empty spacetime \cite{MTW}
\begin{eqnarray}
B_{\alpha\beta\mu\nu}&:=&R_{\alpha\lambda\mu\sigma}R_{\beta}{}^{\lambda}{}_{\nu}{}^{\sigma}
+R_{\alpha\lambda\nu\sigma}R_{\beta}{}^{\lambda}{}_{\mu}{}^{\sigma}
-\frac{1}{8}g_{\alpha\beta}g_{\mu\nu}R_{\lambda\sigma\rho\tau}R^{\lambda\sigma\rho\tau},\\
S_{\alpha\beta\mu\nu}&:=&R_{\alpha\mu\lambda\sigma}R_{\beta\nu}{}^{\lambda\sigma}
+R_{\alpha\nu\lambda\sigma}R_{\beta\mu}{}^{\lambda\sigma}
+\frac{1}{4}g_{\alpha\beta}g_{\mu\nu}R_{\lambda\sigma\rho\tau}R^{\lambda\sigma\rho\tau}.
\end{eqnarray}

\section{Derive tetrad and Einstein superpotentials}
Consider the first order Lagrangian (4-form) density \cite{So}
\begin{equation}
{\cal{}L}:=dq\wedge{}p-\Lambda(q,p),
\end{equation}
where $q$ and $p$ are canonical conjugate variables, $\Lambda$ is
the potential.  The corresponding Hamiltonian 3-form (density) is
\begin{equation}\label{7Apr2007}
H(N):=\pounds_{N}q\wedge{}p-i_{N}{\cal{}L},
\end{equation}
where the Lie derivative $\pounds_{N}:=i_{N}d+di_{N}$.  The
interior product of the Lagrangian is
\begin{equation}
i_{N}{\cal{}L}=\pounds_{N}q\wedge{}p-\epsilon{}i_{N}q\wedge{}dp-\epsilon{}dq\wedge{}i_{N}p
-i_{N}\Lambda-d(i_{N}q\wedge{}p),
\end{equation}
where $\epsilon=(-1)^{f}$ and $q$ is an $f$-form.  Then
(\ref{7Apr2007}) can be rewritten as
\begin{equation}
H(N)=\epsilon{}i_{N}q\wedge{}dp+\epsilon{}dq\wedge{}i_{N}p+i_{N}\Lambda
+d{\cal{}B}(N),
\end{equation}
where the natural boundary term is
\begin{equation}\label{20Feb2007}
{\cal{}B}(N)=i_{N}q\wedge{}p.
\end{equation}
This is called the quasilocal boundary expression, because when
one integrates the Hamiltonian density over a finite region to get
the Hamiltonian, the boundary term leads to an integral over the
boundary of the region.  Let
\begin{equation}\label{20aFeb2007}
q\rightarrow{}-\frac{1}{2\kappa}\eta_{\alpha}{}^{\beta},
\quad\quad{}p\rightarrow \Gamma^{\alpha}{}_{\beta}.
\end{equation}
Rewrite (\ref{20Feb2007})
\begin{equation}
2\kappa{\cal{}B}(N)
=\Gamma^{\alpha}{}_{\beta}\wedge{}i_{N}\eta_{\alpha}{}^{\beta}
=-\frac{1}{2}N^{\alpha}U_{\alpha}{}^{[\mu\nu]}\epsilon_{\mu\nu},\label{17Mar2007}
\end{equation}
where $N^{\alpha}$ is the timelike vector field and the
superpotential
\begin{equation}\label{27Mar2007}
U_{\alpha}{}^{[\mu\nu]}
=-g^{\beta\sigma}\Gamma^{\tau}{}_{\beta\lambda}\delta^{\lambda\mu\nu}_{\tau\sigma\alpha}.
\end{equation}
The connection in (\ref{27Mar2007}) is free for any frames as long
as the superpotential gives the physical sensible results, such as
inside matter (mass density) and at spatially infinity (ADM mass).
It is tetrad \cite{Nester} if the connection is using orthonormal
frames, while Freud \cite{Freud} employed holonomic frames. The
corresponding gravitational energy are studied by \cite{Nester}
and \cite{MTW} respectively.  Similarly, one can use differential
form to derive the boundary expression instead of using the
superpotential.  Furthermore, selectively combining orthonormal
and holonomic frames through the connection can obtain the
desirable gravitational result.  This is the main issue in the
present paper and the detail will be discussed in the next
section.

\section{Combination of tetrad and Einstein expressions}
The pseudotensor can be obtained as
\begin{equation}
t_{\alpha}{}^{\mu}=\partial_{\nu}U_{\alpha}{}^{[\mu\nu]}.
\end{equation}
Taking the second derivatives of this pseudotensor gives the
gravitational energy-momentum density.  Deser $et$ $al$.
\cite{Deser} consider the combination of Einstein
$E_{\alpha\beta}$ and Landau-Lifschitz $L_{\alpha\beta}$
pseudotensors as follows
\begin{equation}
\partial^{2}_{\mu\nu}\left(L_{\alpha\beta}+\frac{1}{2}E_{\alpha\beta}\right)
=B_{\alpha\beta\mu\nu}.
\end{equation}
This combination gives a good result simply because the
Bel-Robinson tensor. Using the differential form, consider the
middle term of (\ref{17Mar2007})
\begin{equation}\label{17aMar2007}
d{\cal{}B}(N)
=\frac{N^{\mu}}{2\kappa}\left(R^{\alpha}{}_{\beta}\wedge\eta_{\alpha}{}^{\beta}{}_{\mu}
-\Gamma^{\alpha}{}_{\beta}\wedge\Gamma^{\lambda}{}_{\alpha}\wedge\eta_{\lambda}{}^{\beta}{}_{\mu}
-\Gamma^{\alpha}{}_{\beta}\wedge\Gamma^{\lambda}{}_{\mu}\wedge\eta_{\alpha}{}^{\beta}{}_{\lambda}
\right).
\end{equation}
Inspired by \cite{Deser}, suppose the connection with the
following combination
\begin{equation}\label{29Mar2007}
\Gamma^{\alpha}{}_{\beta}=\left\{\frac{s}{2}R^{\alpha}{}_{\beta\nu\mu}
-\frac{k}{3}(R^{\alpha}{}_{\beta\mu\nu}+R^{\alpha}{}_{\mu\beta\nu})\right\}x^{\nu}dx^{\mu}
+{\cal{}O}(x^{2}),
\end{equation}
where $s,k$ are real numbers and $s+k=1$ for normalization.
Moreover, the curvature tensor with $s$ refers to orthonormal
frames and $k$ means holonomic frames respectively. Using
(\ref{29Mar2007}), rewrite (\ref{17aMar2007})
\begin{eqnarray}
d{\cal{}B}(N)&=&-\frac{N^{\mu}}{2\kappa}\left\{ 2G^{\rho}{}_{\mu}
+\frac{1}{4}\left[
\begin{array}{cccc}
s^{2}B^{\rho}{}_{\mu\xi\kappa}
+\frac{sk}{3}(5B^{\rho}{}_{\mu\xi\kappa}
-\frac{1}{2}S^{\rho}{}_{\mu\xi\kappa})\\
+\frac{2k^{2}}{9}(4B^{\rho}{}_{\mu\xi\kappa}-S^{\rho}{}_{\mu\xi\kappa})
\quad\quad\quad\quad\\
\end{array}
\right] x^{\xi}x^{\kappa}
\right\}\eta_{\rho}\nonumber\\
&{}&+{\cal{}O}({\rm{}Ricci},x)+{\cal{}O}(x^{3}).
\end{eqnarray}
The first term of this expression is dominated by the Einstein
tensor $G^{\rho}{}_{\mu}$ which means it satisfies the condition
inside matter at the origin because of the equivalence principle.
When it goes to the higher order which indicates the gravitational
energy.  As mentioned before $s+k=1$, consider the three cases.
Case (i): When $k=0$ which is the tetrad teleparalle gauge current
energy-momentum expression $M_{\alpha\beta}$ for pure orthonormal
frames and the associated second derivatives at the origin
\cite{Nester} is
\begin{equation}
\partial^{2}_{\xi\kappa}M^{\rho}{}_{\mu}=\frac{1}{2}B^{\rho}{}_{\mu\xi\kappa}.
\end{equation}
Case (ii): When $k=1$ which is the Einstein pseudotensor and the
corresponding second derivatives \cite{MTW} is
\begin{equation}
\partial^{2}_{\xi\kappa}E^{\rho}{}_{\mu}
=\frac{1}{9}(4B^{\rho}{}_{\mu\xi\kappa}-S^{\rho}{}_{\mu\xi\kappa}).
\end{equation}
Case (iii): When $k=-3$, gravitational energy-momentum density of
this particular selective ratio expression $t_{\alpha\beta}$ is
\begin{equation}
\partial^{2}_{\xi\kappa}t^{\rho}{}_{\mu}=2B^{\rho}{}_{\mu\xi\kappa}.
\end{equation}
This is a good result we found in the present paper.  It only
contains the pure positive Bel-Robinson tensor and the combination
between orthonormal and holonomic frames are unique.

\section{Conclusion}
The energy-momentum for a gravitating system is considered by the
tetard teleparalle gauge current in orthonormal frames; it gives
good result, namely the Bel-Robinson tensor.  Likewise, the
classical Einstein pseudotensor has been used in a holonomic
frames to investigate the same subject, unfortunately it does not
have the desired outcome.  Deser $et$ $al$. used a combination of
the second derivatives of the Einstein and Landau-Lifschitz
pseudotensors to obtain the Bel-Robinson tensor. Inspired by their
work, we found a gravitational energy expression which enjoys the
positive energy property from a combination of the tetrad
expression and the Einstein pseudotensor, meaning that the
connection coefficients are being selected by the uniquely
specific intermediate between orthonormal and holonomic frames.

\section*{Acknowledgment}
This work was supported by NSC 96-2811-M-032-001.

\end{document}